\documentstyle[12pt]{article}

\begin{document}
\newcommand{\cl}{\centerline}
\renewcommand{\theequation}{\arabic{equation}}
\newcommand{\beq}{\begin{equation}}
\newcommand{\eeq}{\end{equation}}
\newcommand{\bea}{\begin{eqnarray}}
\newcommand{\eea}{\end{eqnarray}}
\newcommand{\nn}{\nonumber}
\newcommand\pa{\partial}
\newcommand\un{\underline}
\newcommand\ti{\tilde}
\newcommand\pr{\prime}
\begin{titlepage}
\setlength{\textwidth}{5.0in} \setlength{\textheight}{7.5in}
\setlength{\parskip}{0.0in} \setlength{\baselineskip}{18.2pt}

\begin{flushright}
{\tt gr-qc/0702018} \\
{\tt SOGANG-HEP 318/07}
\end{flushright}

\vspace*{0.3cm}

\begin{center}
{\large\bf Entropy of a charged black hole in two dimensions without
cutoff}
\end{center}

\begin{center}
{Wontae Kim$^{1,a}$, Yong-Wan Kim$^{2,b}$, and Young-Jai
Park$^{1,c}$} \par
\end{center}

%\vskip 0.4cm
\begin{center}
{$^{1}$Department of Physics and Center for Quantum Spacetime,}\par
{Sogang University, Seoul 121-742, Korea}\par \vskip 0.5cm
{$^{2}$National Creative Research Initiative Center for Controlling
Optical Chaos, Pai-Chai University, Daejeon 302-735, Korea}\par
\end{center}
\vskip 0.3cm

\begin{center}
(\today)
\end{center}
\vfill

\vskip 0.4cm

\begin{center}
{\bf ABSTRACT}
\end{center}
\begin{quotation}

By introducing the generalized uncertainty principle (GUP) on
quantum density states, we newly obtain a consistent entropy of a
scalar field on the (1+1)-dimensional Maxwell-dilaton background
without an artificial cutoff in contrast to the results of the brick
wall model, which depend on the choice of the Hawking temperature
for the extremal case.

\vskip 0.4cm


\noindent
---------------------------------------------------------------------\\
\noindent
$^a$wtkim@sogang.ac.kr \\
\noindent $^b$ywkim@pcu.ac.kr \\
\noindent $^c$yjpark@sogang.ac.kr
%\vskip 0.5cm
\noindent

\end{quotation}
\end{titlepage}

\newpage

\section{Introduction}
Three decades ago, Bekenstein suggested that the entropy of a black
hole is proportional to the area of the horizon through the
thermodynamic analogy \cite{bek}. Subsequently, Hawking showed that
the entropy of the Schwarzschild black hole satisfies exactly the
area law by means of Hawking radiation based on the quantum field
theory \cite{haw}. After their works, 't Hooft investigated the
statistical properties of a scalar field outside the horizon of the
Schwarzschild black hole by using the brick wall method (BWM) with
the Heisenberg uncertainty principle (HUP) \cite{tho}. The entropy
proportional to the horizon area is obtained, however, a brick wall
cutoff, which was introduced to remove the divergence of the density
of states, looks unnatural. This method has been used to study the
statistical properties of bosonic and fermionic fields in various
black holes \cite{gm,kkps}, and it is found that the general
expression of the black hole entropy consists of the term, which is
proportional to the area of the horizon, and the logarithmic and
infrared divergent terms.

On the other hand, many efforts have been devoted to the generalized
uncertainty relations \cite{gup}, and its consequences, especially
the finite effect on the density of states near horizon. Recently,
in Refs. \cite{li,liu,kkp}, the authors calculated the entropy of
various black holes, which is proportional to the horizon area
related to the inverse power term of the minimal length, by using
the new equation of state density motivated by the generalized
uncertainty principle (GUP) \cite{gup}, which drastically solves the
ultraviolet divergences of the just vicinity near the horizon
without any cutoff. It is, however, not clear how this proposal with
the GUP works in two dimensions when the horizon is just a point. Up
to now, the statistical entropy \cite{ms} of the 1+1 extremal
charged black hole was only studied by using the BWM \cite{wtk,ws}
with the HUP. Moreover, for the case of a charged black hole, there
exists the extremal case for which two horizons coalesce. There are
two different choices depending on the Hawking temperature for
entropy of the extremal case. One is to take the nonvanishing
arbitrary period of temperature in the Euclidean formalism and the
corresponding entropy may be a nonzero \cite{hhr}. The other is the
usual vanishing Hawking temperature given by the standard
temperature formula and the entropy could be zero.

In this paper, we study the entropy of a scalar field on the
background of the two-dimensional charged black hole based on the
generalized uncertainty principle. First, we briefly recapitulate
the previous results \cite{wtk} for both nonextremal and extremal
cases in the BWM with the HUP, which were obtained by using an
artificial brick wall cutoff and little mass approximation. But, in
this work, we will avoid the difficulty in solving the Klein-Gordon
wave equation by using the quantum statistical method. For the
nonextremal case the entropy is logarithmic divergent whereas the
extremal case gives the vanishing entropy if we take the standard
vanishing Hawking temperature. In view of the microcanonical
ensemble, however, we shall recast the entropy for the extremal case
by counting the number of the states with zero Hawking temperature
by using the BWM. This result for the entropy amounts to the same
logarithmic-divergent entropy with the nonextremal result. Next, we
derive the free energy of a massive scalar field on the background
of the two-dimensional charged black hole directly by using the new
equation of density states motivated by the GUP in the quantum
gravity. Then, we calculate the quantum entropy of the black hole
via the relation between the free energy and the entropy. As a
result, for both nonextremal and extremal cases, we newly obtain the
desired same nonzero entropy without any artificial cutoff and
little mass approximation in contrast to the BWM case with the HUP.

%%%%%%%%%%%%%%%%%%%%%%%%%%%%%%%%%%%%%%%%%%%%%%
\section{Scalar field on 1+1 Maxwell-dilaton Background}
%%%%%%%%%%%%%%%%%%%%%%%%%%%%%%%%%%%%%%%%%%%%%

Let us now start with the two-dimensional Maxwell-dilaton action
induced by the low energy heterotic string theory \cite{mny,NO},
which is given by
\begin{equation}
\label{action} S = \frac{1}{2\pi} \int d^{2}x
\sqrt{-g}e^{-2\phi}[R+4(\bigtriangledown\phi)^2+ 4\Lambda^2-
\frac{1}{4}F^2],
\end{equation}
where $\phi$ is a dilaton field, $\Lambda^2$ is cosmological
constant, and $F$ is a Maxwell field tensor. The equations of motion
under the variations of the metric, dilaton, and gauge fields are
given by as follows:
\begin{eqnarray}
&& R_{\mu\nu}-\frac{1}{2}g_{\mu\nu}R + 2\left[
\nabla_\mu\nabla_\nu\phi+g_{\mu\nu}\left(
(\nabla\phi)^2-\nabla^2\phi-\Lambda^2+\frac{1}{16}F^2\right)\right]=0,
\nonumber\\
&&
R+4\nabla^2\phi-4(\nabla\phi)^2+4\Lambda^2-\frac{1}{4}F^2=0,\nonumber\\
&& \partial_\nu\left(\sqrt{-g}e^{-2\phi}F^{\mu\nu}\right)=0.
\end{eqnarray}
In the Schwarzschild gauge, the metric and field tensors are
assumed to be
\begin{eqnarray}
\label{metric}
ds^2 &=& - f(r) dt^2 +\frac{1}{f(r)} dr^2, \\
F_{rt} &=& F_{rt}(r).
\end{eqnarray}
The equations of motion in this gauge with respect to the metric,
dilaton, and gauge field yield
\begin{eqnarray}
\label{eqm}
0 &=& \partial_{r}(f\partial_{r}\phi) - 2f(\partial_{r}\phi)^2 -\frac{1}{4}(F_{rt})^2 + 2\Lambda^2, \\
0 &=& \partial^{2}_{r}\phi, \\
0 &=& \partial_{r}(F_{rt}e^{-2\phi}).
\end{eqnarray}
Simple static solutions with suitable boundary conditions are known
as follows:
\begin{eqnarray}
\label{sol} \phi(r) &=& \frac{1}{4}\ln 2 - \Lambda r, \\
F_{rt}(r) &=& \sqrt{2} Q e^{-2\Lambda r}, \\
f(r) &=& 1- \frac{M}{\Lambda} e^{-2\Lambda r}+
\frac{Q^2}{4\Lambda^2}e^{-4\Lambda r}.
\end{eqnarray}
There are two coordinate singularities $r_{\pm}$ which correspond to
the positions of the outer event horizon and the inner Cauchy
horizon:
\begin{equation}
\label{horizon} r_{\pm} = \frac{1}{2\Lambda} \ln \left[
\frac{M}{2\Lambda} \pm \sqrt{\left(\frac{M}{2\Lambda}\right)^2 -
\left(\frac{Q}{2\Lambda}\right)^2}\right],
\end{equation}
where the Cosmic censorship leads to the condition $M\geq Q$. The
Hawking temperature is given by the surface gravity method [10]:
\begin{equation}
\label{Htemp} T_{H} = \frac{\Lambda}{2\pi} [1- e^{-2\Lambda(r_{+} -
r_{-})}],
\end{equation}

In this Maxwell-dilaton background, let us first consider a scalar
field with mass $m$ under the background (7)-(9), which satisfies
the Klein-Gordon equation
\begin{equation}
\label{kg} \frac{1}{\sqrt{-g}} \partial_{\mu}( \sqrt{-g} g^{\mu \nu}
\partial_{\nu} \Phi) - m^{2} \Phi = 0.
\end{equation}
Substituting the wave function $\Phi(r, \theta, t) = e^{-i\omega
t}R(r, {\theta})$, we find that this Klein-Gordon equation becomes
\begin{equation}
\label{rtheta0} \frac{d^{2} R}{{dr}^2}  + \frac{1}{f} \frac{df}{dr}
\frac{dR}{dr} + \frac{1}{f} \left(\frac{\omega^{2}}{f} - m^{2}
\right)R = 0.
\end{equation}

By using the Wenzel-Kramers-Brillouin (WKB) approximation \cite{tho}
with $R \sim exp[iS(r,\theta)]$, we have
\begin{equation}
\label{wkb} {p_{r}}^{2} = \frac{1}{f}\left(\frac{\omega^{2}}{f} -
m^{2}\right),
\end{equation}
where
\begin{equation}
\label{mom} p_{r} = \frac{dS}{dr}.
\end{equation}
On the other hand, we also have the square module momentum
\begin{equation}
\label{smom} p^{2} = p_{r}p^{r} = g^{rr}{p_{r}}^{2} = f{p_{r}}^{2} =
\frac{\omega^{2}}{f} - m^{2}
\end{equation}
with the condition $\omega\geq m \sqrt{f}$.

%%%%%%%%%%%%%%%%%%%%%%%%%%%%%%%%%%%%%%%%%%%%%%
\section{Brick Wall Models with HUP}
%%%%%%%%%%%%%%%%%%%%%%%%%%%%%%%%%%%%%%%%%%%%%

According to the BWM with the HUP, let us briefly recapitulate the
previous work \cite{wtk} in the background of the 1+1 dimensional
charged black hole. However, we would like to avoid the difficulty
of solving the wave equation by using the quantum statistical
method. The usual position-momentum uncertainty relation followed by
the HUP is given by
\begin{equation}
{\Delta x} {\Delta p} \ge \frac{\hbar}{2}.
\end{equation}
From now on we take the units $G=c=\hbar=k_{B}\equiv 1$. When
gravity is ignored, the number of quantum states based on the HUP in
the 1+1 dimension is given by
\begin{equation}
\label{dn} dn = \frac{dr dp_{r}}{2\pi}.
\end{equation}
Then, the number of quantum states with energy less than $\omega$ is
given by
\begin{eqnarray}
\label{nqs} n_{O}(\omega) &=& \frac{1}{2\pi}\int dr dp_{r} \nonumber   \\
&=& \frac{1}{\pi}\int^{L}_{r_{+}+\epsilon} dr \frac{1}{\sqrt{f}}
\left(\frac{{\omega}^2}{f}- m^{2}\right)^{\frac{1}{2}}.
\end{eqnarray}
Note that $\epsilon$ and $L$ are ultraviolet and infrared
regulators, respectively.

On the other hand, for the bosonic case the free energy at inverse
temperature $\beta$ is given by
\begin{equation}
\label{def} e^{-\beta F} = \prod_K
                \left[ 1 - e^{-\beta \omega_K} \right]^{-1}~,
\end{equation}
where $K$ represents the set of quantum numbers. By using Eq.
(\ref{nqs}), the free energy can be rewritten as
\begin{eqnarray}
\label{OfreeE0}
 F_{O} &=& \frac{1}{\beta}\sum_K \ln \left[ 1 - e^{-\beta \omega_K} \right]
   ~\approx ~\frac{1}{\beta} \int dn_{O}(\omega) ~\ln
            \left[ 1 - e^{-\beta \omega} \right]  \nonumber   \\
   &=& - \int^{\infty}_{\mu\sqrt{f}} d \omega \frac{n_{O}(\omega)}{e^{\beta\omega} -1} \nonumber  \\
   &=& - \frac{1}{\pi} \int^{L}_{r_{+}+\epsilon} dr \frac{1}{\sqrt{f}}
   \int^{\infty}_{m\sqrt{f}} d\omega
   \frac{\left(\frac{{\omega}^2}{f}- m^{2}\right)^{\frac{1}{2}}}{e^{\beta\omega} -1}.
\end{eqnarray}
Here we have taken the continuum limit in the first line and
integrated by parts in the second line.

Now, let us evaluate the entropy for the scalar field, which can be
obtained from the free energy (\ref{OfreeE0}) at the Hawking
temperature. Then, the entropy is
\begin{eqnarray}
\label{entropy01} S_{O} &=&  \beta^2 \frac{\partial F_O}{\partial
\beta} |_{\beta = \beta_{H}} \nonumber \\
&=& \frac{\beta^{2}}{\pi} \int^{L}_{r_{+}+\epsilon} dr
\frac{1}{\sqrt{f}} \int^{\infty}_{m\sqrt{f}} d\omega
   \frac{\omega e^{\beta \omega}\left(\frac{{\omega}^2}{f}-
m^{2}\right)^{\frac{1}{2}}}{(e^{\beta\omega} -1)^{2}} |_{\beta =
\beta_{H}},
\end{eqnarray}
where $\beta_{H}$ is the inverse Hawking temperature. Note that at
this stage it is difficult to carry out the analytic integral about
$\omega$ because the value of ${m\sqrt{f}}$ varies depending on $r$
in the wide range $(r_{+}+\epsilon, L)$.

For the case of the massless limit, the entropy becomes
\begin{eqnarray}
\label{entropy02} S_{O} &=& \frac{1}{\beta \pi}
\int^{L}_{r_{+}+\epsilon} dr \frac{1}{f}
\int^\infty_0 \frac{e^{x}x^{2}dx}{(e^{x} -1)^2}|_{\beta = \beta_{H}}, \nonumber \\
&=& \frac{2\zeta(2)}{ \pi\beta_{H}} \int^{L}_{r_{+}+\epsilon} dr
\frac{1}{f},
\end{eqnarray}
where $x=\beta \omega$. Finally, we have the entropy as follows
\begin{eqnarray}
\label{entropy05} &S_{O}&= \frac{\pi}{6\Lambda\beta_H}
\frac{1}{(e^{2\Lambda r_+}-e^{2\Lambda r_-})} \times
\nonumber\\
&&\left[ e^{2\Lambda r_+}\ln\left(\frac{e^{2\Lambda L}-e^{2\Lambda
r_+}}{e^{2\Lambda (r_++\epsilon)}-e^{2\Lambda r_+}} \right)-
e^{2\Lambda r_-}\ln\left(\frac{e^{2\Lambda L}-e^{2\Lambda
r_-}}{e^{2\Lambda (r_++\epsilon)}-e^{2\Lambda r_-}} \right)\right].
\end{eqnarray}

Furthermore, neglecting the finite part of the entropy for simple
consideration, one of us obtained that the most dominant degree of
entropy for the non-extremal case is logarithmic divergent as
\begin{equation}
\label{Sne} S_{O} \approx \frac{1}{12} \ln \frac{1}{2\Lambda
\epsilon}
\end{equation}
with the cancelation of the infrared divergent $L$-terms, while, for
the extremal case just on account of $\beta\rightarrow \infty$ as
\begin{equation}
\label{Sext} S^{ext}_{O} = \beta^2 \frac{\partial F_O}{\partial
\beta} |_{\beta \rightarrow \infty} = 0 ~~(r_+ = r_- ).
\end{equation}
On the other hand, for the extremal case from a zero temperature
quantum-mechanical system around the black hole a similar form of
the entropy was obtained as $S^{ext}_{O} \approx \ln
\frac{1}{2\Lambda \epsilon}$.\cite{wtk} However, these strange
results including logarithmic divergence will be drastically changed
due to the GUP effect in the next section.

%%%%%%%%%%%%%%%%%%%%%%%%%%%%%%%%%%%%%%%%%%%%%%%%%%%%%%%%%%%%%%%%%%%%%
\section{Entropy with Generalized Uncertainty Principle}
%%%%%%%%%%%%%%%%%%%%%%%%%%%%%%%%%%%%%%%%%%%%%%%%%%%%%%%%%%%%%%%%%%%%%%%%%

Recently, many efforts have been devoted to the generalized
uncertainty relation \cite{gup} given by
\begin{equation} {\Delta x} {\Delta p} \ge \frac{1}{2}\left(1 +
{\lambda}({\Delta p})^{2}\right).
\end{equation}
Then, since one can easily get ${\Delta x} \geq \sqrt{\lambda}$,
which gives the minimal length, it can be defined to be the
thickness of the thin-layer near horizon, which naturally plays a
role of the brick wall cutoff. Furthermore, based on the generalized
uncertainty relation, the volume of a phase cell in the 1+1
dimensions is changed from $(2{\pi})$ into
\begin{equation}
\label{gup} 2{\pi}(1 + {\lambda}{p^{2}}),
\end{equation}
where $p^2 = p^{r}p_{r}.$

From Eq. (\ref{gup}), the number of quantum states with energy less
than $\omega$ is given by
\begin{eqnarray}
\label{Tnqs} n_{I}(\omega) &=& \frac{1}{2\pi} \int dr dp_{r}
 \frac{1}{1+ {\lambda}(\frac{{\omega}^2}{f}- m^{2})} \nonumber   \\
&=& \frac{1}{\pi}\int dr \frac{1}{\sqrt{f}}
\frac{\left(\frac{{\omega}^2}{f}- m^{2}\right)^{\frac{1}{2}}}{1+
{\lambda}(\frac{{\omega}^2}{f}- m^{2})}.
\end{eqnarray}
Note that it is convergent at the horizon without any artificial
cutoff due to the existence of the suppressing $\lambda$ term in the
denominator induced from the GUP. Then, by using Eq. (\ref{Tnqs}),
the free energy at the Hawking temperature can be obtained as
\begin{eqnarray}
\label{TfreeE}
 F_{I} &=& - \int^{\infty}_{m\sqrt{f}} d\omega \frac{n_{I}(\omega)}{e^{\beta\omega} -1} \nonumber  \\
   &=& - \frac{1}{\pi} \int dr \frac{1}{\sqrt{f}}
   \int^{\infty}_{m\sqrt{f}} d\omega \frac{\left(\frac{{\omega}^2}{f}
   - m^{2}\right)^{\frac{1}{2}}}{(e^{\beta \omega} -1)\left[1+ \lambda
   (\frac{{\omega}^2}{f}-m^2)\right]}.
\end{eqnarray}
From this free energy, the entropy for the massive scalar field is
given by
\begin{eqnarray}
\label{Tentropy0} S_{I} &=& \beta^2 \frac{\partial F_I}{\partial
\beta} |_{\beta =\beta_{H}} \nonumber \\
&=& \frac{\beta^2}{\pi} \int dr \frac{1}{\sqrt{f}}
\int^\infty_{m\sqrt{f}} d\omega \frac{\omega e^{\beta \omega}
\left(\frac{\omega^2}{f}- m^2\right)^{\frac{1}{2}}}
{(e^\beta\omega-1)^2 \left(1+ \lambda (\frac{\omega^2}{f}-
m^{2})\right)} |_{\beta =\beta_{H}}.
\end{eqnarray}
Since $f \rightarrow 0$ near the event horizon, {\it i.e.}, in the
range of $(r_+, r_+ + \epsilon)$, then, without little mass
approximation, the entropy is reduced to

\begin{equation}
\label{Tentropy1} S_{I} = \frac{1}{\pi} \int dr \frac{1}{\sqrt{f}}
\int^\infty_0 dx \frac{f^{-\frac{1}{2}} \beta^{-1}
x^{2}}{(1-e^{-x})(e^{x}
-1)\left(1+\frac{\lambda}{\beta^{2}f}x^{2}\right)}|_{\beta = \beta_{H}}, \nonumber \\
\end{equation}
where $x=\beta \omega$.

On the other hand, we are only interested in the contribution from
the just vicinity near the horizon, $(r_+, r_+ + \epsilon)$, which
corresponds to a proper distance of order of the minimal length,
$\sqrt{\lambda}$. This is because the entropy closes to the upper
bound only in this vicinity, which it is just the vicinity neglected
by BWM. We have
\begin{eqnarray}
\label{invariant} \sqrt{\lambda}=\int_{r_+}^{r_+ +\epsilon}
                          \frac{dr}{\sqrt{f(r)}}
                = \int_{r_+}^{r_+ + \epsilon}
                          \frac{dr}{\sqrt{2\kappa(r-r_{+})}}
                = \sqrt{\frac{2\epsilon}{\kappa}},
\end{eqnarray}
where $\kappa$ is the surface gravity at the horizon of black hole
and it is identified as $\kappa = 2\pi \beta$.

Now, let us rewrite Eq. (\ref{Tentropy1}) as
\begin{equation}
\label{Sdef} S_{I} = \frac{1}{\pi\sqrt{\lambda}} \int^{r_+ +
\epsilon}_{r_{+}} dr \frac{1}{\sqrt{f}} \Sigma_{I},
\end{equation}
where $\Sigma_{I}$ is given by
\begin{equation}
\label{Gdef} \Sigma_{I} = \int^\infty_0 dX
\frac{a^{2}X^{2}}{(e^{\frac{a}{2} X}-e^{- \frac{a}{2}X})^{2}(1+X^2)}
\end{equation}
with $x=\beta\sqrt{\frac{f}{\lambda}}X \equiv aX$. Note that the
entropy is $\beta$ independent at this stage. As $r\rightarrow
r_{+}$, $a$ goes to 0. Since we are interested in the contributions
from just the vicinity of the horizon, the integral equation
(\ref{Gdef}) can be easily solved as follows
\begin{equation}
\label{GI} \Sigma_{I} = \int^\infty_0 \frac{dX}{1+X^2} =
\frac{\pi}{2}.
\end{equation}
Finally, when $r\rightarrow r_{+}$, we newly get a
$\beta$-independent same entropy as follows
\begin{equation}
\label{finalS} S_{I} \approx \frac{1}{\pi\sqrt{\lambda}}\cdot
\sqrt{\lambda} \cdot \frac{\pi}{2} = \frac{1}{2}
\end{equation}
for both nonextremal and extremal black holes. Note that in contrast
to the case of the BWM there is no divergence within the just
vicinity near the horizon due to the effect of the generalized
uncertainty relation on the quantum states.

Now, it seems appropriate to comment on the entropy (\ref{finalS}).
First, for the extremal case ($r_{H}=r_{+}=r_{-}$) we could also
obtain the entropy from a zero temperature quantum-mechanical system
around the black hole with the GUP following the standard formula
for extracting entropy from the partition function such as
\begin{equation}
S^{ext}_{O} = \left(1 -  \beta \frac{\partial}{\partial \beta}
\right)\ln{Z} |_{\beta =\infty}.
\end{equation}
%%%%%%%%%%%
%%%%%%%%%%%%%%%%%%%%%%%%%%%%%%%%%%%%%%%%%%%%%%%%%%%%%%%%%%%%%%%%%%%%%%%%%%%%%%%%%
By using the density of stats $D_{I}(\omega) =
\frac{dn_{I}(\omega)}{d\omega}$, which is obtained from Eq.
(\ref{Tnqs}), the thermodynamic partition function can be obtained
as
\begin{eqnarray}
\label{TfreeE}
 \ln{Z} &=& - \int d\omega D_{I}(\omega) \ln{\left[ 1 - e^{-\beta \omega} \right]} \nonumber   \\
   &=& \frac{1}{\pi} \int^{r_H + \epsilon}_{r_H} dr \frac{1}{\sqrt{f}}
   \int^{\infty}_{m\sqrt{f}} d\omega \frac{\beta \left(\frac{{\omega}^2}{f}
   - m^{2}\right)^{\frac{1}{2}}}{(e^{\beta \omega} -1)\left[1+ \lambda
   (\frac{{\omega}^2}{f}-m^2)\right]}.
\end{eqnarray}
Then, from the above partition function we can easily obtain the
same value of the entropy as that  in Eq. (\ref{finalS}). This means
that by considering the GUP the entropy does not depend on the
choice of Hawking temperature in contrast to the BWM. Second, since
the entropy is proportional to the inverse power term of the minimal
length $\sqrt{\lambda}$, which depends on the $(d+1)$ dimensions as
$S \propto (\sqrt{\lambda})^{-(d-1)}$ \cite{li,liu,kkp}, we have
consistently obtained the $\sqrt{\lambda}$ independent desired
entropy in the (1+1)-dimensions.

In summary, we have investigated the massive scalar field within the
just vicinity near the horizon of a static black hole in the
background of the charged black hole in the 1+1 dimensions by using
the generalized uncertainty principle. In contrast to the previous
result of the BWM \cite{wtk,ws}, we have newly obtained the proper
entropy without any artificial cutoff and little mass approximation.

\section*{Acknowledgments}

We would like to thank M.S. Yoon for useful discussions. This work
is supported by the Science Research Center Program of the Korea
Science and Engineering Foundation through the Center for Quantum
Spacetime of Sogang University with grant number R11-2005-021.

\end{document}